\documentclass[aps,pre,twocolumn,showpacs,superscriptaddress]{revtex4}
\usepackage{graphicx}
\usepackage{amsmath,amssymb,latexsym}
\bibstyle{apsrev.bib}

\newcommand{\beq}{\begin{equation}}
\newcommand{\eeq}{\end{equation}}

\begin{document}

\title{Emergence of Secondary Motifs in Tube-Like Polymers in a Solvent}

\author{Chiara Poletto}
\affiliation{Dipartimento di Fisica, Universit\`{a} di Padova, via
Marzolo 8 I-35131 Padova}
\author{Achille Giacometti}
\affiliation{Dipartimento di Chimica
Fisica, Universit\`{a} di Venezia, Calle Larga S. Marta DD 2137,
I-30123 Venezia, Italy}
\author{Antonio Trovato}
\affiliation{Dipartimento di Fisica, Universit\`{a} di Padova, via
Marzolo 8 I-35131 Padova}
\affiliation{CNISM, Unit\`a di Padova, Via Marzolo 8, I-35131 Padova, Italy}
\author{Jayanth R. Banavar}
\affiliation{Department of Physics, 104
Davey Laboratory, The Pennsylvania State University, University
Park, Pennsylvania 16802 USA}
\author{Amos Maritan}
\affiliation{Dipartimento di Fisica, Universit\`{a} di Padova, via
Marzolo 8 I-35131 Padova}
\affiliation{CNISM, Unit\`a di Padova, Via Marzolo 8, I-35131 Padova, Italy}
\affiliation{Sezione INFN, Universit\`a di Padova, I-35131 Padova, Italy}

\date{\today}

\begin{abstract}
We study the effects of two kinds of interactions in tube-like polymers
and demonstrate that they result in the formation of secondary motifs. The
first has an entropic origin and is a measure of the effective space
available to the solvent. The second arises from solvophobic interactions
of the solvent with the polymers and leads to an energy proportional to
the contact surface between the tube and solvent particles. The solvent
molecules are modeled as hard spheres and the two interactions are
considered separately with the solvent density affecting their relative
strength. In addition to analytical calculations, we present the results
of numerical simulations in order to understand the role played by the
finite length of short polymers and the discrete versus continuum
descriptions of the system in determining the preferred conformation.
\end{abstract}


\maketitle


The study of entropic effects has a venerable history in
physics. Onsager \cite{Onsager49} studied the nature of an
entropically driven isotropic-nematic transition of a fluid
composed of infinitely thin hard rods. The addition of a small
amount of polymers to a colloidal suspension induces an
effective attraction between the colloidal particles which can
lead to flocculation through the Asakura-Oosawa (AO) mechanism
\cite{Asak54Vr76Lik02}. Our focus, in this letter, is on
studying the optimal conformations of a tube subject to compaction
through its interactions with the solvent. Our principal results are that
the tube adopts  
helical and planar conformations with the latter resulting from 
the discreteness of the tube.

Our results are useful for understanding proteins \cite{Finkelstein02, Fersht}.  In spite of their
complexity, proteins fold into a limited number
\cite{Chothia12} of evolutionarily conserved structures
\cite{DentonChothia3} made up of helices and almost planar sheets. 
It has been proposed that the
geometries of protein folds originate from the common
attributes of all proteins that can be encoded in a simple
geometrical model of a flexible tube of non-zero thickness
\cite{Maritan00, Banavar03, Banavar07}. Proteins
occupy the \emph{marginally compact} phase of a tube-like
polymer \cite{Hoang04, Banavar04, Marenduzzo, Lezon06,
Banavar07} which is rich in secondary motifs and has a
relatively low structural degeneracy compared to the generic
compact polymer phase. While it is well recognized that the formation
of secondary structure arises from backbone-backbone hydrogen
bonds \cite{Pauling12,EisenbergPNAS-Poland}, the
solvent is known to play a fundamental role in promoting the
folding process \cite{Finkelstein02}. The importance of the
solvation free energy and of entropy induced interactions has
been increasingly recognized in the last few years
\cite{Harano05, Snir05, Kinoshita06, Hansen-Goos07}.

In a system of hard spheres, the excluded volume constraint can be
enforced by forbidding configurations in which the distance between
pairs of sphere centers is smaller than the sphere diameter. In the
continuum limit, a self interacting tube-like polymer cannot have two
body interactions \cite{Banavar03b}.  The self-avoidance of a flexible
tube of thickness $\Delta$ (the tube radius) can be enforced through a
three-body potential \cite{Gonzales99, Maritan00, Banavar03b} which
forbids conformations in which the radius of a circle through an
arbitrary triplet of points on the axis is smaller than the tube
thickness. A solvophobic polymer in a solvent, such as a protein in
\textit{vivo}, adopts a self-avoiding conformation in which the
contact surface is minimized, an effect that may be captured by considering
\cite{Banavar07} an effective solvent induced (free-)energy
proportional to the contact surface between the tube and the solvent
particles. Furthermore, due to the
discrete nature of the solvent one expects entropy-induced effective
interactions of the AO type \cite{Asak54Vr76Lik02, Harano05, Kinoshita06} as
investigated by Snir and Kamien \cite{Snir05} and more recently by
Hansen-Goos et al. \cite{Hansen-Goos07} for a tube-like polymer in
helical conformations. The AO type of interaction yields an effective
(free-) energy proportional to the volume excluded to the solvent.

It was noted recently by Hansen-Goos et al.
\cite{Hansen-Goos07, Hansen-Goos07b} that Hadwiger's theorem of
integral geometry \cite{Rota}, under rather general conditions,
allows one to write the solvation free energy $F_{sol}$ of the
tube in a solvent as
\begin{eqnarray}
F_{sol} &=& P V_{exc} + \sigma \Sigma_{acc} + k C + \overline k X
\label{eq14}
\end{eqnarray}
where $V_{exc}$, $\Sigma_{acc}$, $C$ and $X$ are the volume excluded to the
solvent, the area accessible to the solvent and integrated mean
and Gaussian curvatures of the accessible surface,
respectively. $P$ and $\sigma$ are the solvent pressure and the
planar surface tension respectively, whereas $k$ and $\overline
k$ represents the two bending rigidities; the properties of the
solvent enters through these coefficients and can hence be
computed separately. The AO approximation amounts to setting
$\sigma=k=\overline k=0$ and assuming an ideal gas solvent
$P=P_{id}$ corresponding to an infinitely diluted system. The
principal aim of our letter is to provide exact expressions for
the excluded volume and the contact surface of a tube-like polymer
in a solvent of hard spheres and to investigate the effects of
these terms both analytically and by Monte Carlo simulations in
determining the optimal conformations of short polymers.

In a continuum description, the axis of a tube is described by
the curve $\mathbf{R}(s)$ parameterized by its arc length $s$.
The Frenet reference frame at location $s$ is given in terms of
the unit vectors
$\{\hat{\mathbf{T}}(s),\hat{\mathbf{N}}(s),\hat{\mathbf{B}}(s)\}$,
the tangent, the normal and the binormal respectively
\cite{CoxeterKamien02}. The curvature at location $s$,
$\kappa(s)$ (the inverse radius of curvature), is defined in
terms of one of the Frenet-Serret equations,
$d\hat{\mathbf{T}}\left(s\right)/d s = \kappa\left(s\right)
\hat{\mathbf{N}}\left(s\right)$ and
$\hat{\mathbf{B}}\left(s\right)= \hat{\mathbf{T}}\left(s\right)
\times \hat{\mathbf{N}}\left(s\right)$.

The position vector of a point inside the tube of length $L$ is
\begin{eqnarray}
\mathbf{r}\left(s,\rho,\theta\right) &=& \mathbf{R}\left(s\right) +
\rho \left[ \cos \theta ~\hat{\mathbf{N}} \left(s\right) + \sin \theta
~\hat{\mathbf{B}}\left(s\right) \right]
\label{eq3}
\end{eqnarray}
where $0 \le \theta \le 2 \pi$, $0 \le \rho \le \Delta $ and $0
\le s \le L$. The volume element in the above curvilinear
coordinate system is given by $dV=  \rho\left \vert
\chi\left(s,\rho,\theta\right) \right\vert  ds d\rho d\theta$,
where $\chi\left(s,\rho,\theta\right) \equiv 1-\kappa(s) \rho
\cos \theta $. Let $\epsilon$ be the radius of the solvent
particles assumed to be spherical. For a straight tube the
excluded volume is simply $V_{exc,S}=\pi L
\Delta_{\epsilon}^2$, where
$\Delta_{\epsilon}\equiv\Delta+\epsilon$, since the volume
inaccessible to the centers of the solvent particles is a tube
with thickness $\Delta_{\epsilon}$. In order to calculate the
excluded volume when the tube curls and different regions of
the tube come close to each other, we have to consider points
which are shared by more than one circular section of radius
$\Delta_{\epsilon}$ associated with distinct points of the tube
axis. We avoid multiple counting by introducing the
integer function $n(s,\rho,\theta)$ which gives the number of
circular cross sections of radius $\Delta_{\epsilon}$, centered
at distinct points of the tube axis, containing the point
$\mathbf{r}(s,\rho,\theta)$ where $\rho<\Delta_{\epsilon}$. A
point, $\mathbf{r}(s,\rho,\theta)$, belongs to the circular
section associated with $\mathbf{R}(s')$ if the vector distance
$\mathbf{d}_{s,\rho,\theta}(s')=\mathbf{r}(s,\rho,\theta)-\mathbf{R}(s')$
is perpendicular to the tangent at $s'$,
$\hat{\mathbf{T}}(s')$, and its modulus is less than
$\Delta_{\epsilon}$. Note that $n(s,\rho,\theta) \ge 1$ because
$s'=s$ trivially satisfies the above conditions. The excluded
volume is thus given by
\begin{eqnarray}
V_{exc} &=& \int_\mathcal{D}\ dV\
\frac{1}{n\left(s,\rho,\theta\right)} \label{eq8}
\end{eqnarray}
where $\mathcal{D}$ is the domain $0 \le \theta \le 2 \pi$, $0
\le \rho \le \Delta_{\epsilon}$ and $0 \le s \le L$. The same
integer function can be used to obtain an expression for the
contact surface on setting the $\rho$ coordinate equal to the
radius $\Delta$ of the tube,
\begin{eqnarray}
\Sigma_{cont} &=& \int_\mathcal{D}\ dV\ \delta(\rho\ -\ \Delta)
\Theta\left[1-n(s,\Delta_{\epsilon},\theta)\right] \label{eq13}
\end{eqnarray}
Here $\Theta(x)$ is the step function equal to $1$ if $x\ge0$ and $0$
otherwise. It is not difficult to show that Eq. \eqref{eq13} is
equivalent to the formula presented in Ref. \cite{Banavar07}, for
both an infinite length tube and a finite one. In the latter case hemispheres
needs to be attached to both ends while computing $n(s,\rho,\theta) \ge
1$, as is the case in our numerical simulations.  In a swollen
configuration corresponding to the absence of overlap,
$n(s,\rho,\theta) = 1$ for all $(s,\rho,\theta)$ and
$\kappa(s)<\Delta_{\epsilon}^{-1}$ for all $s$, and Eqs. \eqref{eq8}
and \eqref{eq13} reproduce the correct results $V_{exc,S}=\pi L
\Delta_{\epsilon}^2$ and $\Sigma_{cont,S}=2 \pi \Delta L$. It is
illuminating to rewrite the difference $V_{exc,S}-V_{exc}$ as
\begin{eqnarray}
\Delta V\equiv  V_{exc,S}-V_{exc}= V_{ov} -2 V_{\chi<0}
\label{eq9}
\end{eqnarray}
where we have introduced the \textit{overlap} volume
\begin{eqnarray}
V_{ov} &=& \int_\mathcal{D}\ dV\
\frac{n\left(s,\rho,\theta\right)-1}{n\left(s,\rho,\theta\right)}
\label{eq10}
\end{eqnarray}
which is the part of the excluded volume with non-vanishing
overlap, i.e. with $n\left(s,\rho,\theta\right)>1$, and the volume
in the highly curved region, i.e. $\kappa(s)>\Delta_{\epsilon}^{-1}$,
\begin{eqnarray}
V_{\chi<0} &=& \int_\mathcal{D}\ dV\
\Theta\left[-\chi\left(s,\rho,\theta\right) \right]\ .
\label{eq11}
\end{eqnarray}
As a result, the difference $\Delta V$ between the excluded
volumes in a swollen and a compact conformation is \textit{not}
simply equal to the overlap volume $V_{ov}$ as in the usual
Asakura-Oosawa mechanism involving rigid spheres
\cite{Asak54Vr76Lik02}. There is no change in shape of the
colloidal particle whereas the tube polymer undergoes a
shape-change upon folding. This difference has some interesting
consequences. Suppose, for instance, that the tube is twisted
and connected at the two ends to form a torus/donut of
thickness $\Delta$ and radius $R$. If $R
> \Delta_{\epsilon}$ one clearly has $\Delta V=0$. If
$R \le \Delta_{\epsilon}$, a straightforward calculation shows
that $V_{ov} =V_{\chi<0}$ so that $\Delta
V=-\Phi(R/\Delta_{\epsilon}) 4 \pi\Delta_{\epsilon}^3/3$
becomes \textit{negative} with the function
\begin{eqnarray}
\Phi\left(x\right) &=&  \left(1+\frac{1}{2} x^2 \right) \sqrt{1-x^2}
-\frac{3}{2} x \arccos x
\label{eq12}
\end{eqnarray}
positive in the interval $0 \le x \le 1$. The swollen configuration is
entropically {\it more} favorable compared to a tight donut. This
counterintuitive result stems from the fact that the curvature yields
both inflated and deflated volume elements. In the absence of overlap,
the two effects balance against each other yielding $\Delta V=0$.
When $\chi<0$, the same balance can be achieved only by allowing
deflated ``negative'' volumes. An imbalanced inflated volume is then
present, which may overturn the excluded volume reduction due to the
presence of overlap, as in the donut case.

\begin{figure}[ht!]
\begin{center}
\includegraphics[angle=-90, scale=0.30]{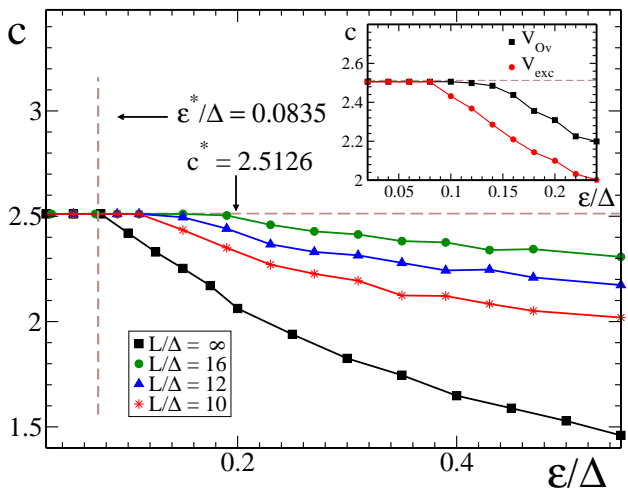} \\
\end{center}
\caption{Plot of the pitch-to-radius value $c=P/R$ for the optimal
helix on minimizing the excluded volume as a function of the ratio
$\epsilon/\Delta$, for different helix lengths, $L$.  Below a given
threshold value (shown for the $L=\infty$ case) of the solvent radius,
the ideal helix conformation is obtained.  Our results were obtained
by first setting the value of the pitch, $P$, so that two different
turns of the helix have minimum distance for a fixed $R$
\cite{Snir05}. Second, we identify the value of $c$ which minimizes
$V_{exc}/L$ for a given $\epsilon$ and these values are plotted
against each other. An analogous calculation for $\Sigma_{cont}/L$
shows a similar trend towards optimal finite helices having a greater
$c$ with respect to the infinite helix. In the inset, the different
behaviors obtained on considering the minimization of the excluded
volume and the maximization of the overlap volume is shown.}
\label{fig1}
\end{figure}

Consider an infinite helix of radius $R$ and pitch $P$. The
translational invariance along the tube axis simplifies the
calculations, and we can consider both the excluded volume and the
contact surface per unit length. Fig. \ref{fig1} shows the values of
$c=P/R$ for the helices which minimizes $V_{exc}/L$, as the solvent
molecule radius changes -- the analogous results for $\Sigma_{cont}/L$
exhibits similar behavior. When $\epsilon$ is below a threshold value,
the optimal configuration corresponds to the \textit{ideal helix}
\cite{Maritan00, Przybyl01, Banavar04, Snir05, Hansen-Goos07} which
\textit{simultaneously} minimizes the local and non-local radius of
curvature \cite{Gonzales99}. For an ideal helix, $c= c^{*}=2.5126\ldots$
which is within $3\%$ of the value for $\alpha-$helices in proteins \cite{Maritan00}.
Interestingly, a similar optimality condition is also found for double
strand DNA \cite{Stasiak00}. These observations suggest that
$\alpha-$helices might originate in general optimization processes
transcending the large differences in the amino acid sequences of
different proteins. As the solvent sphere size $\epsilon$ increases
beyond the threshold, Fig. \ref{fig1} shows that the optimal value of
$c$ decreases from its value in the ideal helical structure. Our
results confirm those obtained in previous work \cite{Snir05,
Hansen-Goos07}. For the the excluded volume interaction, the helix
maintains its ideal pitch-to-radius ratio $c^{*}$ up to a critical
value $\epsilon^{*} \simeq 0.0835 \Delta$. Above this threshold, the
ideal helix is no longer the optimal conformation -- even though the
excluded volume along the central axis increases, there is a
better overlap between subsequent turns thus decreasing the excluded
volume. The inset of Fig. \ref{fig1} shows the results maximizing the
overlap rather than minimizing the excluded volume. The optimal value
$c^{*}$ is maintained up to $\epsilon ^{*}\simeq 0.1 \Delta$, a value
larger than that obtained in the minimum excluded volume case. This
mirrors the situation with the donut: for large curvature, there is a
non-zero, albeit small, difference between $\Delta V$ and $V_{ov}$ as
dictated by Eq. \eqref{eq9}.
An exact analytical computation for the contact surface
\cite{Trovato07} yields an identical pattern with a smaller critical value
$\epsilon^{*}= 0.04627 \Delta$ for the solvent radius. Again, this is in
perfect agreement with the value given in Ref. \cite{Hansen-Goos07} using
a different method. Our method allows one to treat the case of
\textit{finite} helices efficiently using numerical integration.
Fig. \ref{fig1} shows the optimal value of $c$ for helices of varying
length, $L$. Short helices show a tendency to remain ideal for higher
values of the solvent radius.

When the tube is very long or there are many independent tubes, one
might expect that the helical conformation is destabilized.  For many
long tubes, an optimal configuration which minimizes both the contact
surface and the excluded volume is likely a regular arrangement of
parallel tubes forming a hexagonal arrangement somewhat reminiscent of
the Abrikosov phase in Type II superconductors \cite{Tinkham96}. We
have verified that $N \ge 4$ infinitely long tubes have a lower
excluded volume per unit length with respect to the optimal helix (for a
single tube one should also take into account the $N-1$ turns).

\begin{figure}[ht!]
\begin{center}
\includegraphics[scale=0.40]{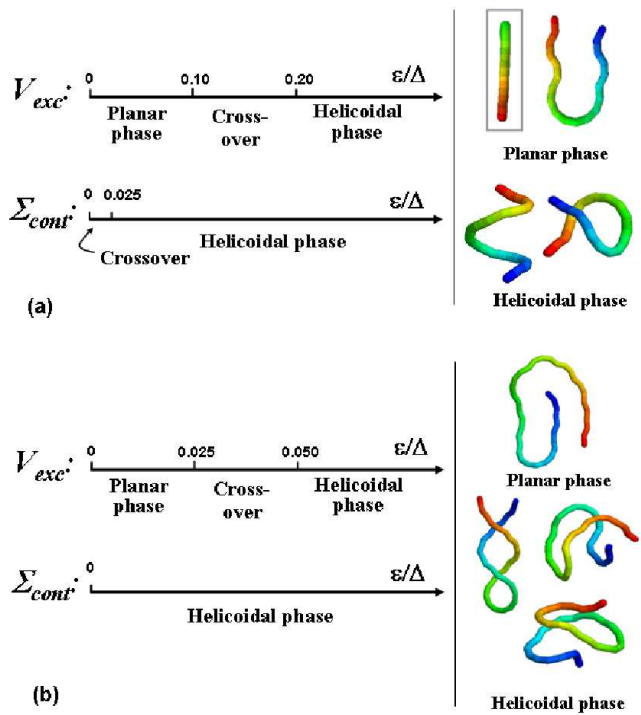}
\end{center}
\caption{Conformations adopted by tubes of length (a) $L=10\Delta$ and (b)
$L=20\Delta$ subject to either the excluded volume or contact surface
prescriptions for promoting compaction. The values of
$\epsilon/\Delta$ are 0.005, 0.025, 0.05, 0.10, 0.20 and 0.50. The
resulting conformations for the shorter tube are either saddles or helices
for $\epsilon/\Delta\gtrsim 0.20$, while for smaller $\epsilon$, after a
crossover phase, the hairpin becomes the ground state -- the planarity of
the structure is highlighted on the left of (a). For the longer tube, the
planar phase consist of $\beta$-sheet-like structures and the helicoidal
phase is characterized by double helices, saddles and irregular
helices. The planar phase disappears for the contact surface case.
All these conformations are akin to those found in Ref. \cite{Banavar04}.}
\label{fig2}
\end{figure}

\begin{table}[!tbh]
  \begin{center}
    \begin{tabular}{c|cc|cc}
      \hline
      \hline
      &
      \multicolumn{2}{|c|} {$L=10 \Delta$} &
      \multicolumn{2}{|c} {$L=20 \Delta$} \\
      $\epsilon/\Delta$ & $\,\Delta V/(L \Delta^2)\,$ & $\,\Delta \Sigma/(L \Delta)\,$ & $\,\Delta V/(L \Delta^2)\,$ & $\,\Delta \Sigma/(L \Delta)\,$\\

      \hline
      0.005 & 0.049 & 0.62 & 0.053 & 0.98 \\
      0.025 & 0.061 & 0.80 & 0.067 & 1.2  \\
      0.050 & 0.079 & 0.98 & 0.098 & 1.5  \\
      0.100 & 0.12  & 1.2  & 0.18  & 1.9  \\
      0.200 & 0.24  & 1.6  & 0.39  & 2.1  \\
      0.500 & 0.77  & 2.1  & 1.3   & 2.9  \\
      \hline
      \hline
    \end{tabular}
  \end{center}
  \caption{Values of $\Delta V/(L \Delta^2)$ and  $\Delta \Sigma/(L \Delta )$ related to the optimal structures, for the tested values of $\epsilon/\Delta$ and $L/\Delta$. } 
\label{tab:1}
\end{table}

In order to investigate the possibility for other kinds of optimal
conformations for a short tube as well as the role of discrete versus
continuum descriptions of the tube, we resort to Monte Carlo
simulations using the formulae given above for both the excluded
volume and the contact surface. We seek to identify the conformations
which minimize $V_{exc}$ or $\Sigma_{cont}$ as the solvent radius
$\epsilon$ and the polymer length $L$ are varied (the thickness
$\Delta$ is kept fixed because the results depend only on the
dimensionless ratios $\epsilon/\Delta$ and $L/\Delta$). Our Monte Carlo
simulations combine crankshaft and pivot moves of a homopolymer with a
Metropolis annealing schedule \cite{annealing} following initializing
the chain in an arbitrary extended conformation. We have considered two
different polymer lengths, $L=10\Delta$ and $L=20\Delta$, fixing the
bond length $b=\Delta/2$.  The results are summarized in Fig. \ref{fig2}
-- the optimal structures obtained on varying $\epsilon$ are shown -- while 
the values of $\Delta V= V_{exc,S}-V_{exc}$ and $\Delta \Sigma= \Sigma_{cont,S}-\Sigma_{cont}$ reached are 
reported in Tab. \ref{tab:1}. Fig. \ref{fig2} (a) displays the conformations 
obtained with $L=10\Delta$. For both the excluded volume and
contact surface, two different ``phases" emerge: at small {\bf
$\epsilon$,} a \emph{planar phase}, in which the ground state has a
typical planar hairpin structure, is found; at large {\bf $\epsilon$,} a
\emph{helicoidal phase} is found in which helices and saddles dominate,
with comparable $V_{exc}$ and $\Sigma_{cont}$. The two different regimes
are separated by a \emph{crossover} region characterized by the
coexistence of all these structures. Note that the excluded volume effect
exhibits a stronger propensity for the planar phase. This is evident for
$L=20\Delta$ shown in Fig.\ref{fig2} (b) -- the contact surface
planar phase disappears already at $\epsilon/\Delta = 0.005$, the lowest
value tested. The helicoidal phase in this case consist of double helices,
saddles and irregular helices, with turns of different lengths. Our
simulations suggest the existence of an energy barrier between the two
classes of conformations (hairpin versus helix or saddle) in the crossover
region. The appearance of planar structures is a consequence of the
discrete nature of the polymer (non-zero bond length)  which plays a
crucial role especially at small $\epsilon$.

In summary, tube-like polymers with naturally arising solvent induced
interactions exhibit low free energy conformations with secondary
motifs. This suggests that the secondary motifs commonly found in
bio-polymers such as proteins and DNA have a common and fundamental
origin which transcends chemical detail. When the discrete nature of
the polymer dominates, planar conformations akin to the $\beta$-sheet
in proteins emerge along with helical conformations as the optimal
ones. In contrast, when the solvent molecule size is sufficiently
large so that the discrete nature of the polymer can be neglected, the
continuum approximation is valid and helical conformations dominate
with single and double helix conformations.

This work was supported by PRIN no. 2005027330 in 2005 and INFN.


\end{document}